# Nonlinear Phenomena In Ferromagnetic Double Layers

M.A. Malugina, Yu.P. Sharaevsky

*Saratov State University, Saratov, Russia*

The non-stationary nonlinear models of magnetostatic waves propagation in layered ferromagnetic structures are developed. This models are based on use of the coupled nonlinear Schrodinger equations for amplitude of a bending around taking into account an electrodynamic coupling between layers. The basic nonlinear processes in such structures, in particular, the effects of self-modulation and solitons formation are studied numerically, in comparison with the similar effects in single ferromagnetic films.

**1.** To the present time nonlinear self-influence processes (self-modulation, self-focusing, formation of envelope solitons etc.) in one-film ferromagnetic structures are investigated in detail, both theoretically and experimentally for different types of magnetostatic waves (MSW) [1]. The wide possibilities for research of the indicated nonlinear effects present multilayer ferromagnetic structures, which can include as layers of a different physical nature [2], so and several ferromagnetic films [3-4]. The linear wave processes in multilayer ferromagnetic structures are investigated enough. The coupling influence on nonlinear wave processes in multilayer ferromagnetic structures can result in a modification not only a dispersion low of magnetostatic waves, but also nonlinearity of a system.

The present report is devoted to a numerical modeling of self-influence processes, which is accompanied by a modulation instability, as applied to structure consisting of two ferromagnetic layers. We study the features of these nonlinear processes in comparison with similar in single films. The special attention is given to a problem, which is connected with a pulse propagation in similar structure and possibility of envelope «solitons» formation. The propagation of continuous signals in similar structures, being accompanied with a self-modulation of a signal, is also considered for different type of excitation of system on distances there is a lot of smaller length of linear swapping of energy between layers.

**2.** The analysis was conducted with reference to ferromagnetic structure consisting of two identical ferromagnetic layers by the width $D$, with a saturation magnetization $M_0$, separated by a dielectric layer of width $d$. The considered structure is infinite in $x$ and $y$ direction, the constant magnetic field $H_0$ is enclosed on a normal in a surface of layers, that corresponds to distribution of forward volume MSW in ferromagnetic layers (FVMSW). The coefficient of coupling between layers is determined as $exp[-kd]$. In a layered structure the dispersing curve for a single film, is decomposed on two, appropriate to normal modes of a coupled system - fast and slow. In a linear approximation because of small angles of deflection of a vector of a magnetization from an equilibrium position it is reasonable to assume: $M_{z_{1,2}} \approx M_0$, where $M_{z_{1,2}}$ - is longitudinal components of a vector of a magnetization. The nonlinearity of MSW is accounted by a variation of longitudinal components of magnetic moment, for which it is possible to note: $M_{z_{1,2}} \approx M_0 \left(1 - \left(M_{x_{1,2}}^2 + M_{y_{1,2}}^2\right)/2M_0^2\right)$ Where $M_{x_{1,2}}$, $M_{y_{1,2}}$ - is a transversal components of a magnetization for appropriate ferromagnetic layers, what dependent on amplitudes of a fast and slow waves. Let's present wave functions for fast and slow normal waves as:

$\phi_{f,sl}(y,t) = \varphi_{f,sl}(y,t) e^{j\psi_{f,sl}}$, where $\varphi_{f,sl}(y,t)$ - slowly varying complex envelope amplitudes, $\psi_{f,sl}$ - phases of normal waves.

In assumption of the electrodynamics coupling of a high-frequency fields in a coupled films we obtained the system of nonlinear wave equations describing the evolution of complex envelope amplitudes of the normal waves. This system, with account of a dissipation, can be presented as:

$$\begin{cases} j\left(\dfrac{\partial \varphi_f}{\partial t} + V_f \dfrac{\partial \varphi_f}{\partial y}\right) + A_f \dfrac{\partial^2 \varphi_f}{\partial y^2} - \dfrac{B_f}{4}\left(|\varphi_f|^2 + |\varphi_{sl}|^2\right)\varphi_f + j\alpha\varphi_f = 0 \\ j\left(\dfrac{\partial \varphi_{sl}}{\partial t} + V_{sl} \dfrac{\partial \varphi_{sl}}{\partial y}\right) + A_{sl} \dfrac{\partial^2 \varphi_{sl}}{\partial y^2} - \dfrac{B_{sl}}{4}\left(|\varphi_f|^2 + |\varphi_{sl}|^2\right)\varphi_{sl} + j\alpha\varphi_{sl} = 0. \end{cases} \quad (1)$$

In the equations (1) $V_{f,sl}$, $A_{f,sl}$, $B_{f,sl}$ - is a group velocity, dispersion and nonlinearity of normal waves, $\alpha$ - is a factor of a dissipation. The variations of group velocity and dispersion depending on magnitude of connection are shown on fig.1, influence of connection on coefficients of nonlinearity is not significant.

On fig.1 it is possible to allocate areas (I, II, III, IV) of a typical variation of the coefficients. In the areas I and III coefficients of a dispersion of both normal waves are negative, in the areas II, IV – they have a opposite signs, besides has place a distinction in group velocities of normal waves between areas.

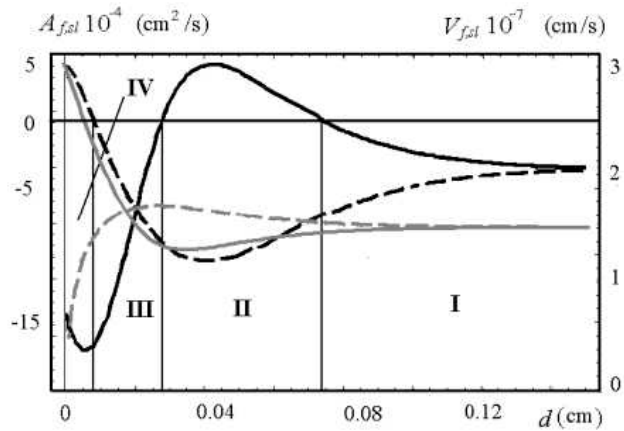

Fig.1. Dependence of dispersion (dark curves) and group velocity (light) fast (continuous curves) and slow (dashed curves) normal waves from distance between films.

**3.** The main results obtained on the basis of a numerical solution of a system of equations (1) are reduced to the following:
- When is injected signals of equal amplitude in coupled films (in a phase or antiphase) the wave in a system can be both stable, or unstable of rather small longitudinal envelope perturbations; in a single film FVMSW is always unstable. The given effect is connected with a changing of a sign of normal modes dispersion (fig.1). Since for given type of excitation, the system (1) is degenerated in 2 disconnected equations for each of normal modes; the Lighthill criterion is not satisfied, for example, in the areas II and IV, in which FVMSW in a coupled system are stable.
- The nonlinear nature of coupling in structure is reveals itself in following: at excitement of one of the films, the signal of large amplitude is divided between coupled films, the signal of small amplitude is not redistributed between films (fig.2). The division of a signal has a place on length of structure, which is inversely proportional to power of an input signal and depends on distance between films.

Under pulsed excitement of structure:
- When is injected signal in one of the coupled films of structure a velocity and amplitude of soliton-like formation on receive antenna is more, than in a single line and increases in accordance with the films convergence in structure (increasing of coupling). That is connected with change the velocities of normal waves depending on coupling (fig.1) and simultaneous influence of losses.
- At excitation of one film in coupled structure the splitting of an input pulse is observed, as against a single line. Thus, using electrodynamics coupling between magnetostatic waves, is possible to control processes of a soliton-like pulse formation in multilayer ferromagnetic structures.

At excitement of system by continuous signal:
- Under inphase or antiphase excitation of coupled structure a decreasing of «threshold» value of an input amplitude, necessary for existence of a self-modulation in system, in comparison with appropriate «threshold» value without of coupling, has a place.
- The nonlinear character of coupling between layers of structure is reveals itself in increasing of frequency of a self-modulation, in comparison with a single layer; the influence of coupling in this case is similar to an influence of nonlinearity. Besides, the variation of coupling between films can result in appearance of additional frequencies and in a transition to chaos of pulse envelope amplitude in case of initial excitement of only one of the films. Also, it is possible a return to a stationary mode with further variations of distance between film. The similar behavior of a signal in layered structure is connected with two normal waves simultaneously existing in structure, with a different dispersion, group velocity and nonlinearity, and, therefore, a frequency of a self-modulation. It is possible to modify the chaos area depending on the film parameters (thickness, magnetisation, losses and ext.) and the input pulse characteristics (frequency, effective power, ext.), what can to be of interests for formation of various types of a signals in the microwave transmission lines.

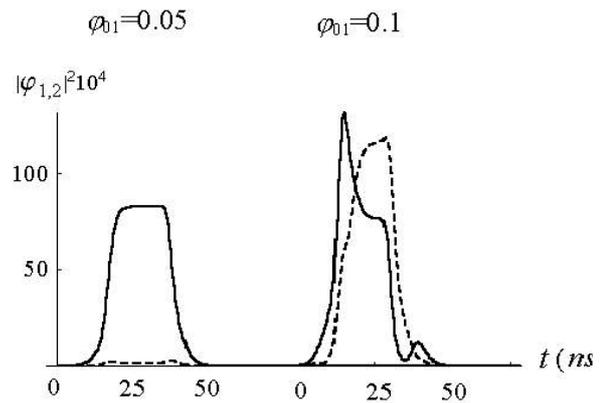

Fig.2. The form of pulse envelope in a film 1 (continuous curve) and film 2 (dotted) of a coupled system at distance $y = 0.2$ cm from an output antenna for different input amplitudes ($\varphi_{01} = 2\varphi_0$, $\varphi_{02} = 0$).